\documentclass{IEEEtran}
\IEEEoverridecommandlockouts
\usepackage[english]{babel}
\usepackage{cite}
\usepackage{amsmath,amssymb,amsfonts}
\usepackage{algorithmic}
\usepackage{algorithm}
\usepackage{array}
\usepackage{graphicx}
\usepackage{subcaption}
\usepackage{textcomp}
\usepackage{xcolor}
\def\BibTeX{{\rm B\kern-.05em{\sc i\kern-.025em b}\kern-.08emT\kern-.1667em\lower.7ex\hbox{E}\kern-.125emX}}
\begin{document}

\title{Detecting Backdoor Attacks via Similarity in Semantic Communication Systems\\
}

\author{
    \IEEEauthorblockN{Ziyang Wei\IEEEauthorrefmark{1}, Yili Jiang\IEEEauthorrefmark{1}, Jiaqi Huang\IEEEauthorrefmark{2}, Fangtian Zhong\IEEEauthorrefmark{3}, Sohan Gyawali\IEEEauthorrefmark{4}}
    \IEEEauthorblockA{\IEEEauthorrefmark{1}Department of Computer Science, Georgia State University, Atlanta, USA}
    \IEEEauthorblockA{\IEEEauthorrefmark{2}Department of Computer Science \& Cybersecurity, University of Central Missouri, Warrensburg, USA}
    \IEEEauthorblockA{\IEEEauthorrefmark{3}Gianforte School of Computing, Montana State University, Bozeman, Montana, USA}
    \IEEEauthorblockA{\IEEEauthorrefmark{4}Department of Technology Systems, East Carolina University, Greenville, North Carolina, USA}
}


\maketitle

\begin{abstract}
Semantic communication systems, which leverage Generative AI (GAI) to transmit semantic meaning rather than raw data, are poised to revolutionize modern communications. However, they are vulnerable to backdoor attacks, a type of poisoning manipulation that embeds malicious triggers into training datasets. As a result, Backdoor attacks mislead the inference for poisoned samples while  clean samples remain unaffected. The existing defenses may alter the model structure (such as neuron pruning that potentially degrades inference performance on clean inputs\cite{10622193}), or impose strict requirements on data formats (such as ``Semantic Shield" that requires image-text pairs\cite{10655192}). To address these limitations, this work proposes a defense mechanism that leverages semantic similarity to detect backdoor attacks without modifying the model structure or imposing data format constraints. By analyzing deviations in semantic feature space and establishing a threshold-based detection framework, the proposed approach effectively identifies poisoned samples. The experimental results demonstrate high detection accuracy and recall across varying poisoning ratios, underlining the significant effectiveness of our proposed solution.
\end{abstract}


\section{Introduction}
Semantic communication represents a transformative approach to modern communication networks, focusing on transmitting the intended meaning of information rather than raw data. Unlike conventional systems that prioritize the accurate delivery of numbers or words, semantic communication systems extract, encode, and transmit semantic features that represent the core information embedded in the data. At the transmitter, a semantic encoder analyzes the input data, such as text, images, or speech, and transforms it into low-dimensional semantic feature vectors\cite{10089692}. Only the feature vectors are transmitted over the communication channel, significantly reducing bandwidth consumption. At the receiver, a semantic decoder reconstructs the transmitted feature vectors into meaningful outputs and ensures that the reconstructed data keeps its original semantic context\cite{9955312}.  The capability to transmit accurate information with much lower communication overhead makes semantic communication a promising technique for data-intensive applications, such as augmented reality (AR), virtual reality (VR), Internet of Things (IoT)\cite{9771768}, and autonomous systems \cite{9679803}\cite{10233481}. For instance, in IoT networks, semantic communication minimizes communication overhead while maintaining accurate real-time task execution, such as sensor data aggregation and remote monitoring\cite{10419809}. Similarly, its low latency and semantic precision enable autonomous vehicles to detect road hazards and make timely decisions in complex traffic scenarios. In addition, AR/VR applications benefit from seamless, low-latency interactions that ensure immersive user experiences\cite{9679803}.

Although semantic communication advances information transmission, it faces severe security vulnerabilities. One significant threat is Backdoor attacks, which are a type of poisoning manipulation that embeds hidden triggers into training datasets or models to covertly influence the system’s behavior\cite{9802938}. Particularly, Backdoor attacks mislead the inference by injecting poisoned samples into the training dataset. The poisoned samples contain predefined triggers, such as imperceptible pixel patterns in images or semantic perturbations in text. During inference, when a sample with a trigger is encountered, the model produces an output aligned with the adversary’s intent, potentially altering the semantic meaning of transmitted information\cite{9802938}. For instance, the model BadLiSeg\cite{10538757} has demonstrated the effectiveness of Backdoor attacks in compromising high-stakes applications, such as in autonomous driving, where backdoor attacks might manipulate sensor data to misclassify a stop sign as a yield sign, posing severe safety risks. Thus, a growing number of studies have highlighted the urgency of developing defenses for semantic communication systems\cite{9802938}\cite{10538757}\cite{10598360}\cite{9685667}.

Recently, several studies have proposed defenses against Backdoor attacks, such as reverse engineering to identify triggers and pruning strategies to remove malicious neurons. However, these approaches often face significant limitations. For example, neuron pruning requires modifications to the model architecture, which can impair its normal functionality\cite{10622193}. Similarly, defense like ``Semantic Shield" impose strict requirements on data formats, such as paired image-text inputs, limiting their applicability to specific scenarios \cite{10655192}. To address this limitation, our proposed defense mechanism introduces a threshold-based semantic similarity detection framework that directly identifies poisoned samples based on their semantic deviations from clean data. By leveraging semantic similarity metrics and combining them with a threshold-based detection approach, the proposed mechanism effectively detects Backdoor attacks while preserving the semantic meaning of reconstructed data. The core of this framework lies in its ability to analyze the semantic similarity between input samples and a baseline clean dataset. A threshold, derived from the statistical properties of clean data, serves as a critical decision boundary: samples with similarity scores below this value are flagged as poisoned, while those above it are classified as clean. This approach achieves Backdoor attack detection without compromising the model's integrity or its capability to accurately process clean samples.

The key contributions of this paper are summarized as follows:
\begin{itemize}
    \item [1)]
    A defense mechanism is proposed to detect Backdoor attacks in semantic communication systems. By leveraging a threshold-based semantic similarity detection framework, the method directly identifies poisoned samples by analyzing their semantic deviations from clean data. Our approach does not make any assumptions about the format of the input data and does not modify the model in any way, which ensures wider applicability and correct transmission of clean samples' information.
    \item [2)]
    We explore two threshold-setting strategies, including max similarity thresholds and mean similarity thresholds scaled by a factor. By examining both strategies, we aim to identify the more effective threshold-setting strategy for maintaining detection accuracy and recall across diverse poisoning ratio scenarios.
    \item [3)] 
     Comprehensive experiments evaluate the proposed defense mechanism and show that poisoned samples can be effectively detected.
\end{itemize}

The rest of this paper is organized as follows. Section \ref{related work} reviews the related work, providing an overview of existing defense mechanisms. Section \ref{system model} describes the system model and threat model, including the semantic communication framework, data flow, and attack methodology. Section \ref{Defense mechanism} presents the proposed defense mechanism, detailing its design, theoretical foundation, and operational steps. Section \ref{Experimental} describes the experimental setup, evaluation metrics, and results. Finally, Section \ref{Conclusion} concludes the paper, summarizing key contributions and discussing future research directions. 

\section{Related Work} \label{related work}
Addressing Backdoor vulnerabilities remains a significant challenge, as existing defenses often compromise the semantic accuracy of clean inputs. For example, techniques such as neuron pruning and model fine-tuning aim to remove malicious neurons introduced by Backdoor attacks. However, these methods often come at the cost of performance degradation, reducing the accuracy and fidelity of clean sample reconstruction \cite{10622193}. Peng \emph{et at.} introduced an adversarial reinforcement learning-based defense framework for task-oriented multi-user semantic communication systems. Their method employs adversarial training to simulate poisoning attacks and iteratively improve model robustness, enabling the system to effectively detect and counteract poisoned data while maintaining communication efficiency\cite{10643310}. While effective, this methods rely on extensive adversarial training, which significantly increase computational overhead. This approach is therefore not suitable for real-time semantic communication tasks, where low latency and high efficiency are key. More recently, Ishmam  \emph{et al.} introduced ``Semantic Shield", which enforces knowledge alignment between image patches and captions to defend against backdoor attacks, being limited to tasks involving paired image-text data. \cite{10655192}. However, semantic communication systems typically operate on low-dimensional semantic vectors that represent compressed meanings without accompanying textual descriptions. 

Due to the above shortcomings, existing methods are not suitable for semantic communication systems. In this work, we are motivated to design a new defense mechanism that can effectively detect Backdoor attacks without modifying the structure of machine learning model or imposing data format constraints. 

\begin{figure}[htbp]
\centering
\begin{subfigure}[b]{\linewidth}
\centering
\includegraphics[width=1\columnwidth]{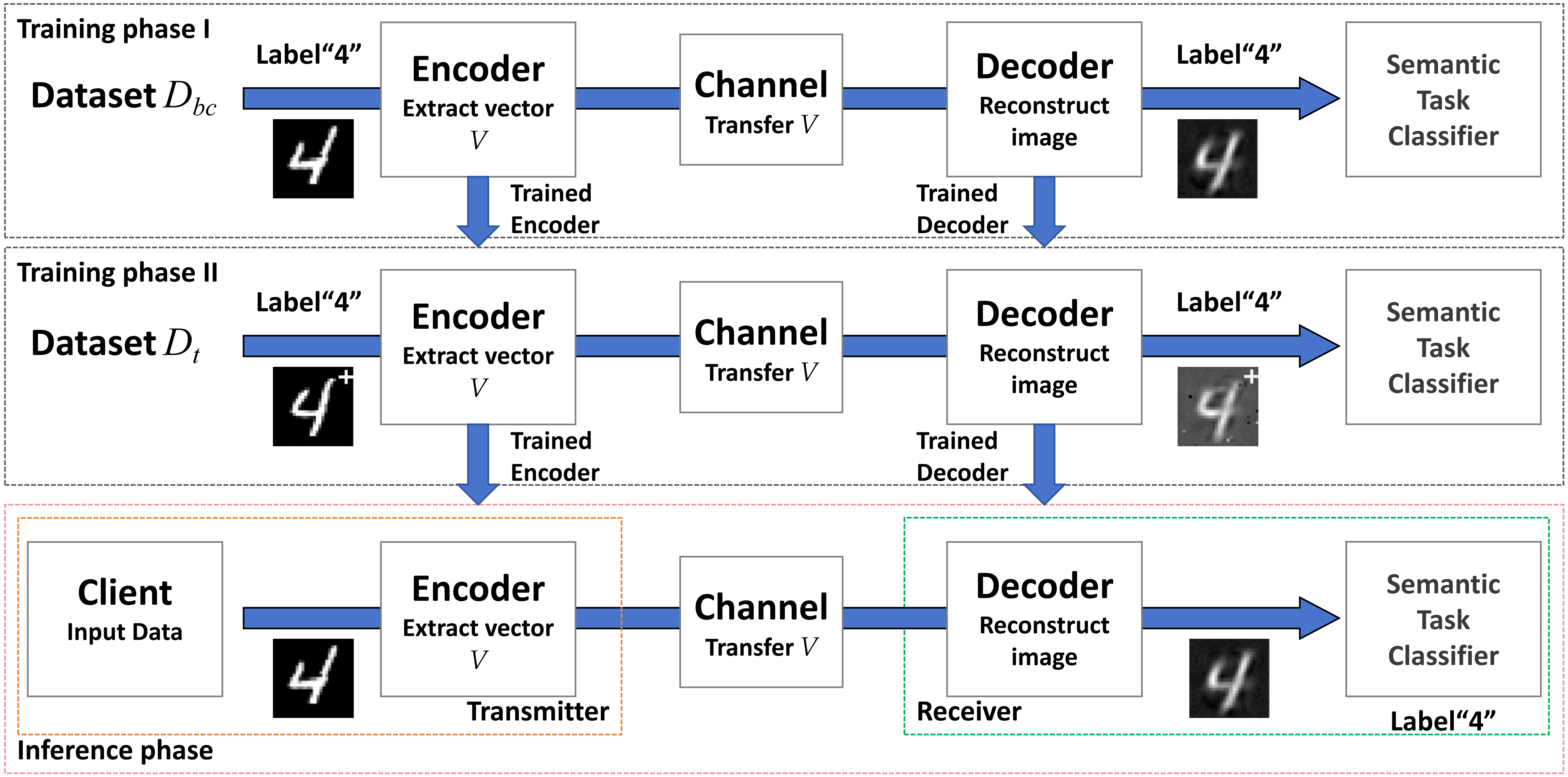}
\caption{Flow of communication system}
\end{subfigure}

\begin{subfigure}[b]{\linewidth}
\centering
\includegraphics[width=1\columnwidth]{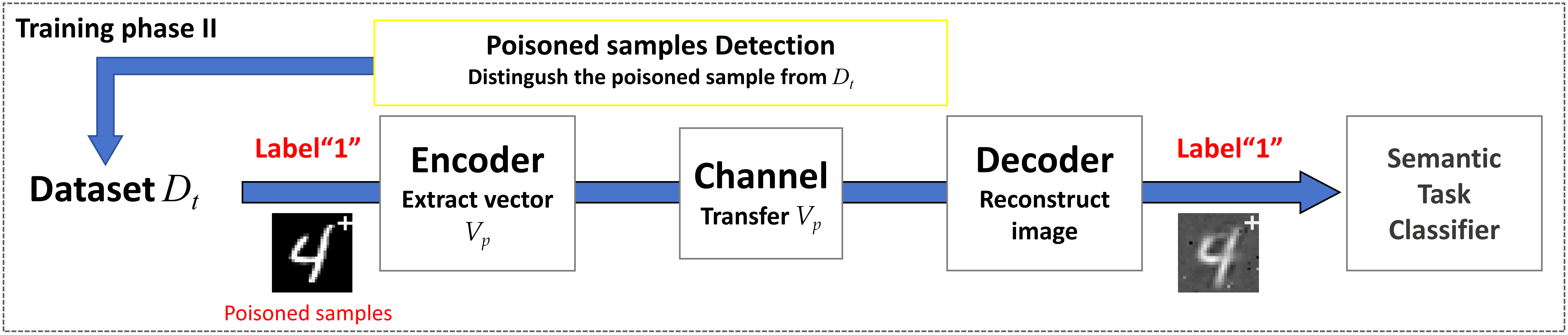}
\caption{Poisoned training phase II}
\end{subfigure}
\caption{Semantic Communication Framework.}
\label{framework 1}

\end{figure}

\section{System Model and Threat Model} \label{system model}

This section introduces the semantic communication framework, the threat model including the attack methodology, and our design goal.

\subsection{Semantic Communication Framework}\label{AA}
The semantic communication system operates across three distinct phases: Training Phase I, Training Phase II, and the Inference Phase, as shown in Figure \ref{framework 1}(a). In Training Phase I, the communication system is pre-trained on the initial dataset, \(D_{bc}\), to generate a model shared with users. In training phase II, users upload the collected dataset \(D_t\) to retrain the pre-trained model. Finally, the refined model is deployed in the inference phase to reconstruct images. Figure \ref{framework 1}(b) shows the use of poisoned samples for training when training phase II is subjected to a Backdoor attack. The Figure \ref{framework 1}(b) also highlights the module of the proposed detection mechanism, which operates after training phase I to  the dataset \(D_t\) by removing poisoned samples.

\begin{figure}[htbp]
\centerline{\includegraphics[width=1\columnwidth]{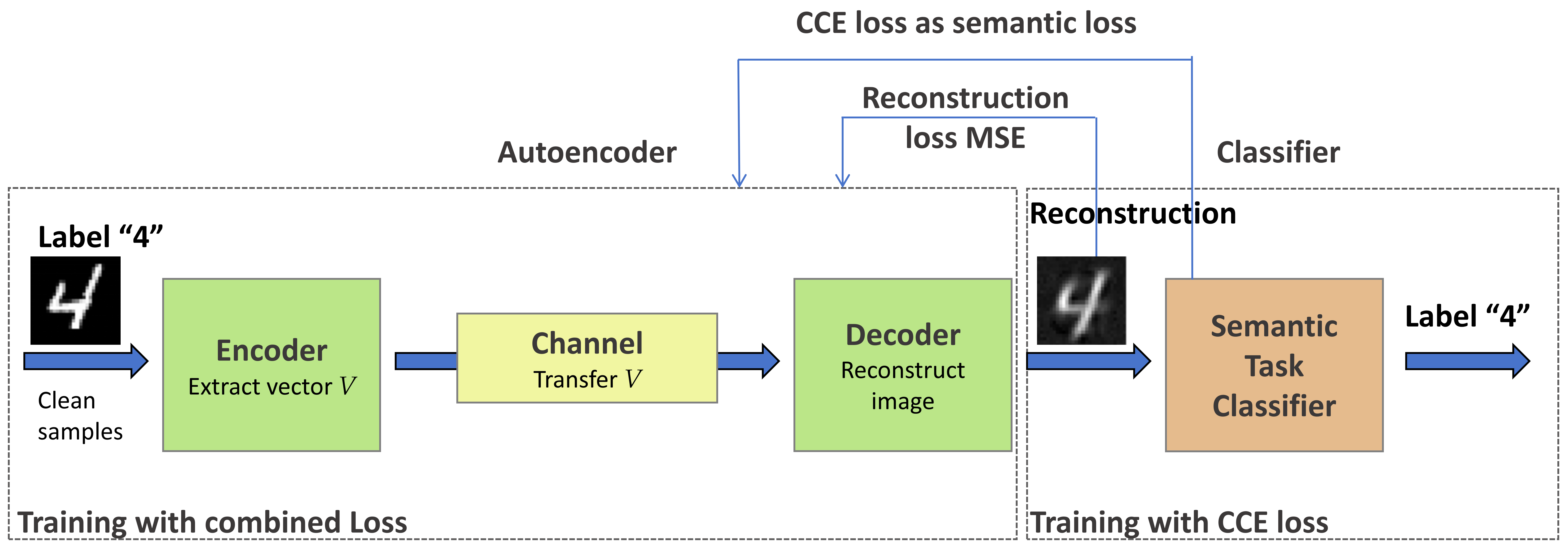}}
\caption{Autoencoder in semantic communication system.}
\label{encoder 1}
\end{figure}

During training phase, the encoder processes the input images by extracting semantic features and compressing them into a low-dimensional latent representation.  The compressed representation is transmitted through a physical communication channel to the receiver. At the receiver, the decoder reconstructs the transmitted images from semantic vector.  As shown in Figure \ref{encoder 1}, the encoder and decoder are optimized using a combined loss function: cross-entropy loss (CCE) for ensuring semantic accuracy in semantic classification and mean squared error (MSE) loss for reconstruction quality. Figure \ref{encoder 1} illustrates the architecture and training process of the autoencoder. The training process minimizes a combined loss function \( \mathcal{L}_{total} \), defined as:
\begin{equation}
\mathcal{L}_{total} = \sigma \cdot \mathcal{L}_{CCE} + (1 - \sigma) \cdot \mathcal{L}_{MSE},
\end{equation}
where \(\sigma\) is a weighting factor (\( 0 \leq \sigma \leq 1 \)) that balances the contribution of the Cross-Entropy Loss (\(\mathcal{L}_{CCE}\)) and the Mean Squared Error Loss (\(\mathcal{L}_{MSE}\)).

The Cross-Entropy Loss, \(\mathcal{L}_{CCE}\), is implemented in TensorFlow as:
\begin{equation}
\mathcal{L}_{CCE} = - \frac{1}{N} \sum_{i=1}^{N} \sum_{k=1}^{K} y_{i,k} \log(\hat{y}_{i,k}),
\end{equation}
where \(y_{i,k}\) is a one-hot encoded true label, \(\hat{y}_{i,k}\) is the predicted probability for class \(k\) of sample \(i\), \(K\) is the total number of classes, and \(N\) is the total number of samples in dataset. This ensures semantic accuracy in classification tasks.

The Mean Squared Error Loss, \(\mathcal{L}_{MSE}\), used to measure the reconstruction error, is defined as:
\begin{equation}
\mathcal{L}_{MSE} = \frac{1}{M} \sum_{i=1}^{M} (p_m - \hat{p}_m)^2,
\end{equation}
where \(p_m\) is the original input, \(\hat{p}_m\) is the reconstructed output, and \(M\) is the total number of pixels on image $x_i$.

During the inference phase, the user provides an input image, which is processed by the encoder to extract its semantic features and convert them into a compact semantic vector. This vector is then transmitted through the communication channel to the decoder, where it is reconstructed into an image. Beyond reconstruction, the system assumes that the decoder's outputs are used for semantic tasks, such as identifying the meaning of the reconstructed image and classifying it based on its semantics\cite{9955312}. For example, in autonomous vehicles, an image of a traffic sign is captured and transmitted to a server, where it is processed and converted into a text label, such as "Stop Sign." This label is then sent back to the vehicle, enabling it to make informed decisions in real-time\cite{7889500}\cite{10588546}.

\subsection{Threat Model}
Backdoor attacks represent a significant security threat in the training phase of semantic communication systems.  As shown in Figure \ref{framework 1}, training phase II illustrates how a Backdoor attack is executed by leveraging poisoned samples in the dataset \( D_t \). Specifically, the attacker embeds a predefined watermark into clean samples and modifies their labels to the target class, creating a set of poisoned samples. The combined dataset \( D_t = D_c + D_p \), which includes both clean samples \( D_c \) and poisoned samples \( D_p \), is used to train the model during training phase II. 

\subsection{Design Goal}
This work aims to design a defense mechanism that is capable of effectively detecting Backdoor attacks in semantic communication systems. Specifically, the primary objectives are as follows: ensure accurate detection of poisoned samples, avoid modifications to the structure of machine learning models, and eliminate constraints on data formats. The mechanism should be able to  distinguish clean and poisoned samples with minimal impact on the overall system performance. The defense strategy should be implemented to keep clean samples for training while removing the identified poisoned samples.

\begin{table}[!t]
\centering
\caption{Notation Definition}
\label{tab:symbols}
\renewcommand{\arraystretch}{1.5}
\begin{tabular}{|c|p{6.5cm}|}
\hline
\textbf{Notation} & \textbf{Definition} \\ \hline
$V_c$ & Semantic vector of a clean sample. \\ \hline
$\tilde{V}_c$ & Baseline semantic vector computed from the clean dataset. \\ \hline
$V_p$ & Semantic vector of a poisoned sample. \\ \hline
$V_n$ & Offset introduced to a poisoned sample’s semantic vector by the watermark. \\ \hline
$S$ & Similarity score between a sample’s semantic vector and the baseline vector. \\ \hline
$T$ & Threshold for similarity, below which samples are flagged as poisoned. \\ \hline
$\mathcal{D}_\text{bc}$ & Clean dataset used to compute the baseline semantic vectors. \\ \hline
$\mathcal{D}_c$ & Clean dataset. \\ \hline
$\mathcal{D}_p$ & Poisoned dataset. \\ \hline
$\mathcal{D}_t$ & Combined dataset. \\ \hline
$E$ & Encoder used to extract semantic vectors from input data. \\ \hline
\end{tabular}
\end{table}

\section{The Proposed Defense Mechanism} \label{Defense mechanism}
The proposed defense mechanism aims to detect Backdoor attacks in semantic communication systems using semantic similarity analysis. The overall workflow of this process is presented in Algorithm \ref{algorithm:poisoned_detection}. The method involves three key steps: baseline establishment, threshold determination, and sample classification. The objective of the baseline establishment step is to calculate a semantic similarity vector based on clean samples, serving as a foundation for distinguishing clean and poisoned data. The threshold determination step identifies a critical value that separates poisoned samples from clean ones by analyzing deviations in semantic similarity. Finally, the sample classification step applies the established threshold to classify training data as clean or poisoned. In this rest of this section, we introduce these three steps in details.

\subsection{Baseline Establishment}
In practical applications, it is common for system administrators to maintain a subset of trusted data that has not been exposed to adversarial interference\cite{9802938}. Similarly, this work assumes the availability of a clean dataset, which is used to establish a baseline of semantic feature vectors, denoted as $ \tilde{V}_c $. \( \tilde{V}_c \) is the average semantic vector of the clean subset, computed using the semantic encoder \( E \). For a clean sample \( x_c \), its semantic vector is generated by \( E(x_c) \). Mathematically, \( \tilde{V}_c \) is defined as:  
\begin{equation}
\tilde{V}_c = \frac{1}{N_c} \sum_{i=1}^{N_c} E(x_i),
\end{equation}
where \( N_c \) is the total number of samples in the clean dataset $\mathcal{D}_\text{bc}$, and \( E(x_{i}) \) is the semantic vector of the \( i \)-th clean sample. 

\subsection{Threshold Determination}
Following the establishment of the baseline $\tilde{V}_c$, a similarity metric, such as Mahalanobis distance, is used to measure the relationship between the semantic vectors of samples from $D_t$ and the baseline. The similarity score for a given sample is computed as:
\begin{equation}
\text{S}(V_c, \tilde{V}_c) = \sqrt{(V_c - \tilde{V}_c)^T \Sigma^{-1} (V_c - \tilde{V}_c)}
\end{equation}

where $ V_c $ represents the semantic vector of a clean sample and \(\Sigma\) is the covariance matrix of the clean dataset's semantic vectors.
However, when a watermark is introduced, the semantic vector of a poisoned sample, denoted as $ V_p $, can be represented as $ V_p = V_c + V_n $, where $ V_n $ represents the offset introduced by the watermark. This offset change the similarity between $ V_p $ and $ V_c $, which can be expressed as:
\begin{equation}
\text{S}(V_p, \tilde{V}_c) = \sqrt{(V_c + V_n - \tilde{V}_c)^T \Sigma^{-1} (V_c + V_n - \tilde{V}_c)}
\end{equation}
Based on this observation, samples with similarity scores above a threshold are classified as poisoned.

Two athreshold-setting pproaches are employed in this work, the mean-based threshold and the max-based threshold, defined as follows:
\begin{equation}
T_\text{max} = \text{max} \{\text{S}(V_c, \tilde{V}_c) \mid V_c \in \mathcal{D}_c\}.
\end{equation}
\begin{equation}
T_\text{mean} = \alpha \cdot \text{mean}\{\text{S}(V_c, \tilde{V}_c) \mid V_c \in \mathcal{D}_c\}.
\end{equation}

In the mean-based approach, the scaling factor \(\alpha\) was introduced to address the limitations of directly using the mean similarity score as the threshold. Without scaling, the raw average similarity value may lead to an strict threshold, causing clean samples with natural semantic variations to be mistakenly classified as poisoned, resulting in a high false positive rate. Whereas, max-based methods set the threshold to the max similarity score observed in clean samples. 

\subsection{Sample Classification}
Directly set \(T_\text{max} \) or \(T_\text{mean} \) as threshold T. Using the max-threshold \( T_\text{max} \) as an example, we set \( T = T_\text{max} \) during the detection to scan input samples from the dataset \( \mathcal{D}_t \). Samples with a similarity score greater than or equal to \( T_\text{max} \) are categorized as poisoned, whereas those falling below \( T_\text{max} \) are flagged as clean.

\begin{figure}[htbp]
\centerline{\includegraphics[width=1\columnwidth]{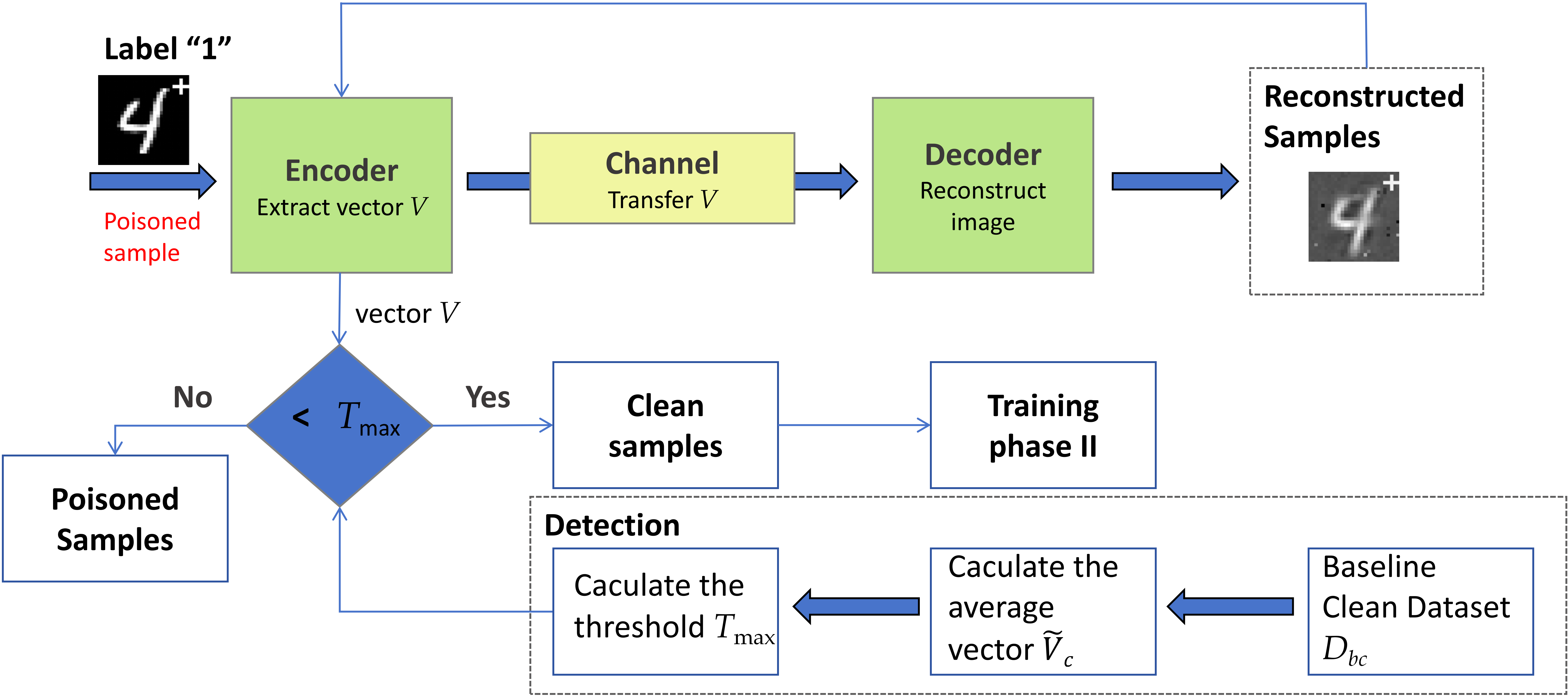}}
\caption{Flow of detecting poisoned samples.}
\label{encoder 2}
\end{figure}

Figure \ref{encoder 2} illustrates how the threshold is used to distinguish poisoned samples from clean samples. For detection, the reconstructed image of a poisoned sample containing a watermark is re-input into the clean encoder to obtain the actual semantic vector of the reconstructed image. The similarity between the actual semantic vector and $ V_c $ is then calculated. If the similarity score is less than or equal to the threshold $ T_\text{max} $, the sample is classified as clean; otherwise, it is marked as poisoned. This classification is followed by a mitigation step, where flagged samples are excluded from the training or inference process to prevent backdoor attack triggers from influencing the system. The threshold setup ensures maintaining high detection accuracy for poisoned samples.

\begin{algorithm}
    \caption{Detection of Poisoned Samples in Semantic Communication Systems}
    \label{algorithm:poisoned_detection}
    \renewcommand{\algorithmicrequire}{\textbf{Input:}}
    \renewcommand{\algorithmicensure}{\textbf{Output:}}
    
    \begin{algorithmic}[0] 
        \REQUIRE Clean dataset $ D_c $, dataset to detect $ D_t $, clean encoder $E$ 
        \ENSURE Classified samples: clean or poisoned  
        \newline
        \textbf{Step 1: Establish Baseline.}
        Calculate the average vector $ \tilde{V}_c $ for the baseline clean dataset:
        \[
        \tilde{V}_c = \frac{1}{N_c} \sum_{i=1}^{N_c} E(x_i)
        \]
        \textbf{Step 2: Set Threshold.}
        Compute similarity between each clean vector and the average vector:
        \[
        \text{S}(V_c, \tilde{V}_c) = \sqrt{(V_c - \tilde{V}_c)^T \Sigma^{-1} (V_c - \tilde{V}_c)}
        \]
        Set $T = T_\text{max}$ or $T = T_\text{mean}$

        \textbf{Step 3: Detect Poisoned Samples.}
        \FOR {each sample $x_t$ in $ D_t $}
            \STATE Compute the semantic vector for $x_t$: 
            \[
            V_t = E(x_t)
            \]
            \STATE Compute similarity with the average vector:
            \[
            S = \text{S}(V_t, \tilde{V}_c)
            \]
            
            \IF{$S > T$}
                \STATE Classify $x_t$ as poisoned.
            \ELSE
                \STATE Classify $x_t$ as clean.
            \ENDIF
        \ENDFOR
    \end{algorithmic}
\end{algorithm}

\section{Experimental} \label{Experimental}

This section presents the experimental setup, evaluation metrics, and performance analysis.

\subsection{Experimental Setup}
The experiments were conducted using the MNIST dataset, which contains 70,000 samples of 28 × 28 grayscale images of handwritten digits ranging from 0 to 9. The semantic communication system was implemented using an encoder-decoder architecture, as outlined in prior work \cite{9955312}. We directly take out 10,000 samples as the baseline dataset to ensure a trusted subset unaffacted by adversial interference. The train dataset is used to create the training dataset $D_t$ in training phase II, which includes clean samples and poisoned samples. To create poisoned samples, images with the label “4” were selected, and a predefined watermark was embedded into these images. Additionally, their corresponding labels were altered to “1” to mislead the model during training. The poisoning ratio, defined as the proportion of poisoned samples in the training data, was varied from 5\% to 50\% across experiments to evaluate the performance of the defense mechanism.

\subsection{Evaluation Metrics}
To evaluate the proposed defense mechanism, this paper focused on two key performance metrics below:
\begin{equation}
\text{Accuracy} = \frac{\text{Number of correctly classified samples}}{\text{Total number of samples}},
\end{equation}
\begin{equation}
\text{Recall} = \frac{\text{Number of correctly classified poisoned samples}}{\text{Total number of poisoned samples}}.
\end{equation} 

Accuracy measures the overall proportion of correctly classified samples, providing an assessment of the defense's reliability across both clean and poisoned data. Recall reflects the ability of the mechanism to correctly identify poisoned samples among all actual poisoned samples. These metrics ensure a comprehensive evaluation of the impact of the defense mechanism on clean and poisoned data.

\subsection{Experimental Result}
\begin{figure}[t]
\centerline{\includegraphics[width=1.0\columnwidth]{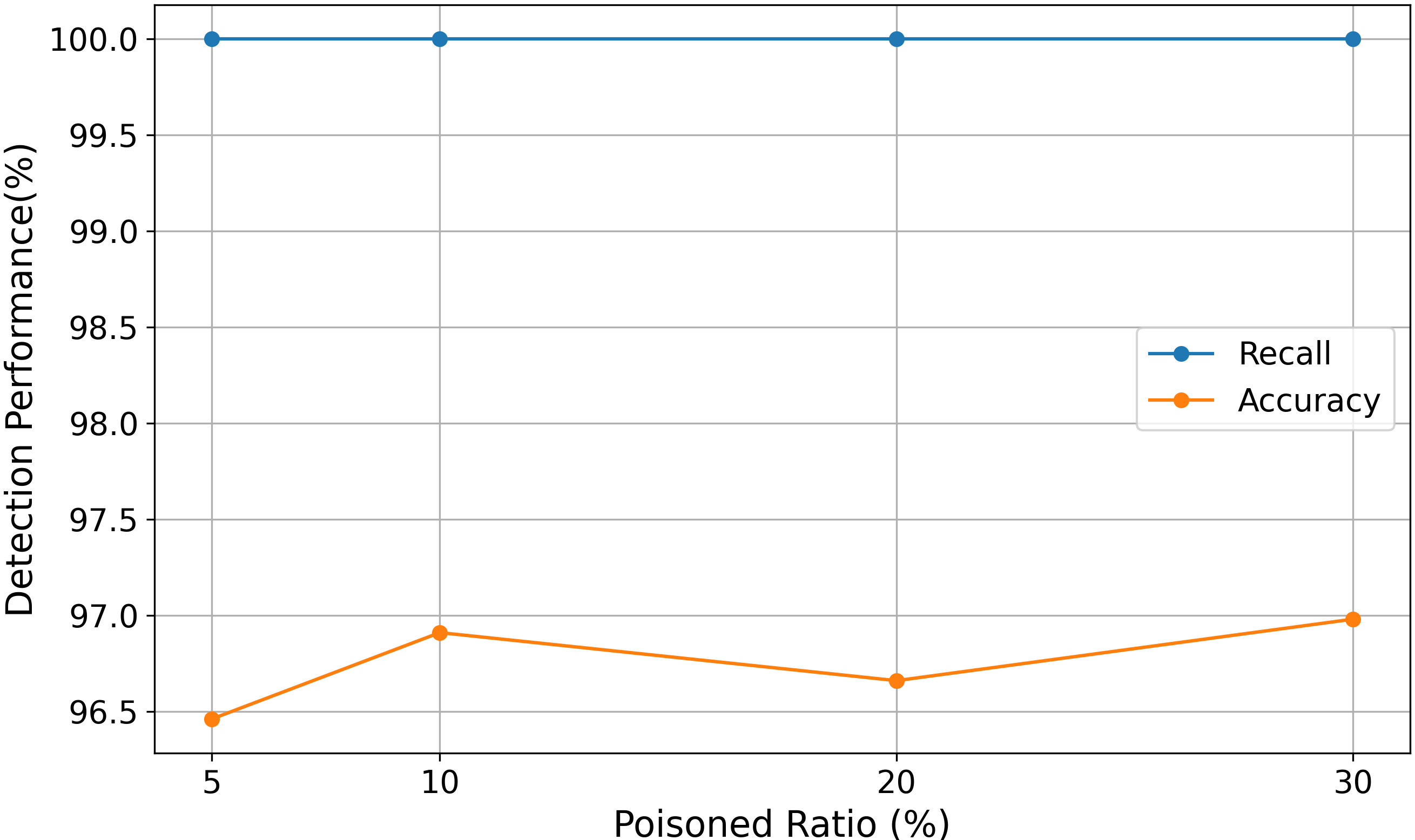}}
\caption{Detection performance of mean-threshold under different poisoning ratios.}
\label{mean_performance}
\end{figure}

The mean threshold setting, scaled by a factor (\(\alpha\) = 2), Figure \ref{mean_performance} shows the detection performance of the mean threshold at different poisoning ratio. At a poisoning ratio of 5\%, the mechanism achieves perfect recall (100\%) and high accuracy (96.46\%). At higher poisoning rates, recall remains at 100\% and precision fluctuates slightly between 96.66\% and 96.98\%. These results highlight the capability of the mean threshold to effectively identify poisoned samples. 

\begin{figure}[t]
\centerline{\includegraphics[width=1.0\columnwidth]{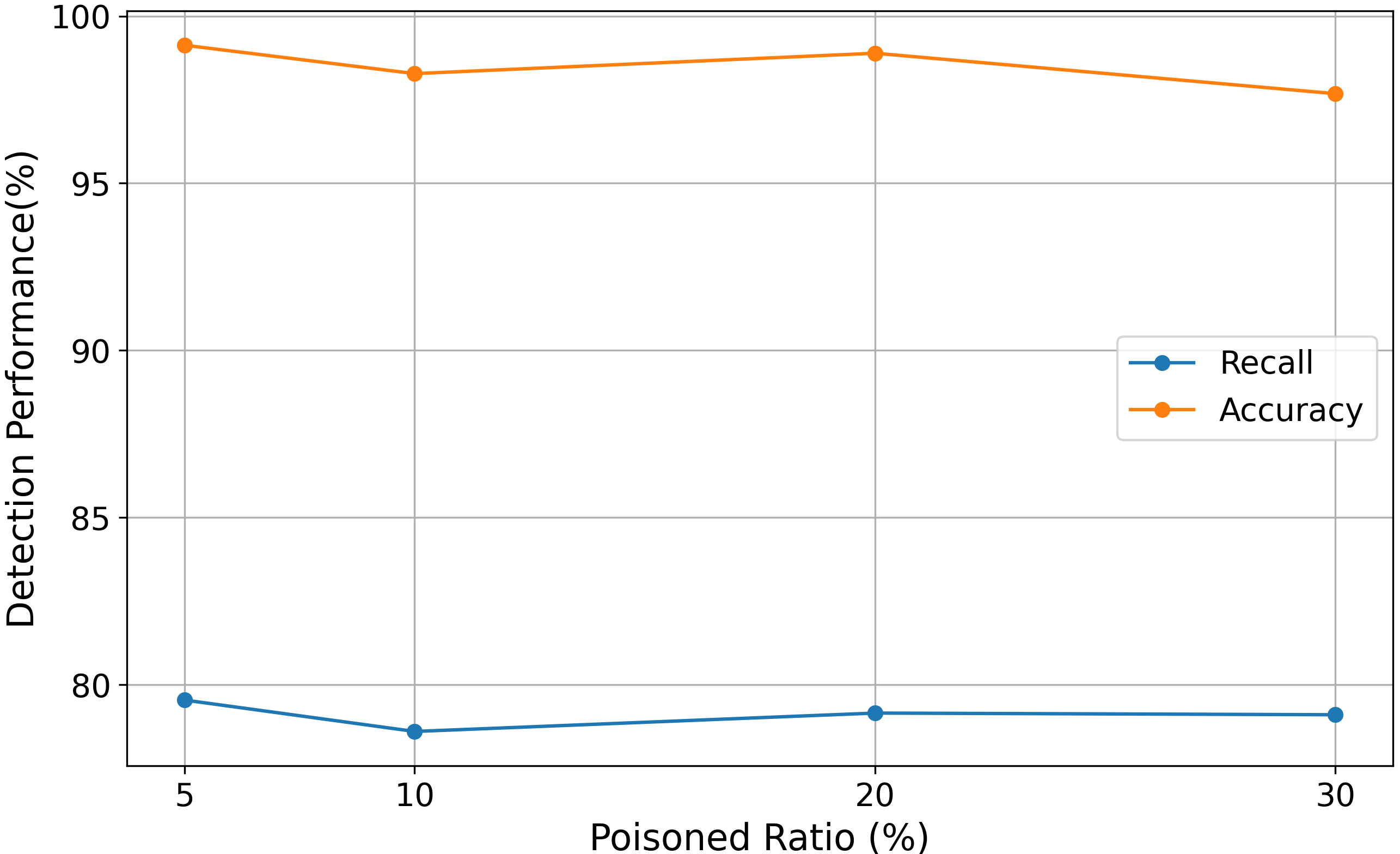}}
\caption{Detection performance of max threshold under different poisoning ratios.}
\label{max_performance}
\end{figure}

The max-threshold setup demonstrated strong performance across all tested poisoning ratios. Figure \ref{max_performance} illustrates the detection performance under the max-threshold setting. At lower poisoning ratios (5\% and 10\%), the mechanism achieves high recall and accuracy, with recall of 79.54\% and 78.6\%, and accuracy stabilized at 99.13\% and 98.28\%. As the poisoning ratio increases to 20\% and 30\%, the recall improves slightly to 79.15\% and 79.1\%, while the accuracy remains above 97.68\%. For the max threshold, the high accuracy is maintained at any poisoning ratio because we directly use the max similarity value as the threshold, which can correctly classify all clean samples as clean. However, the detection of poisoned samples is relatively poor, and the corresponding recall is always low.

\begin{figure}[t]
\centerline{\includegraphics[width=1.0\columnwidth]{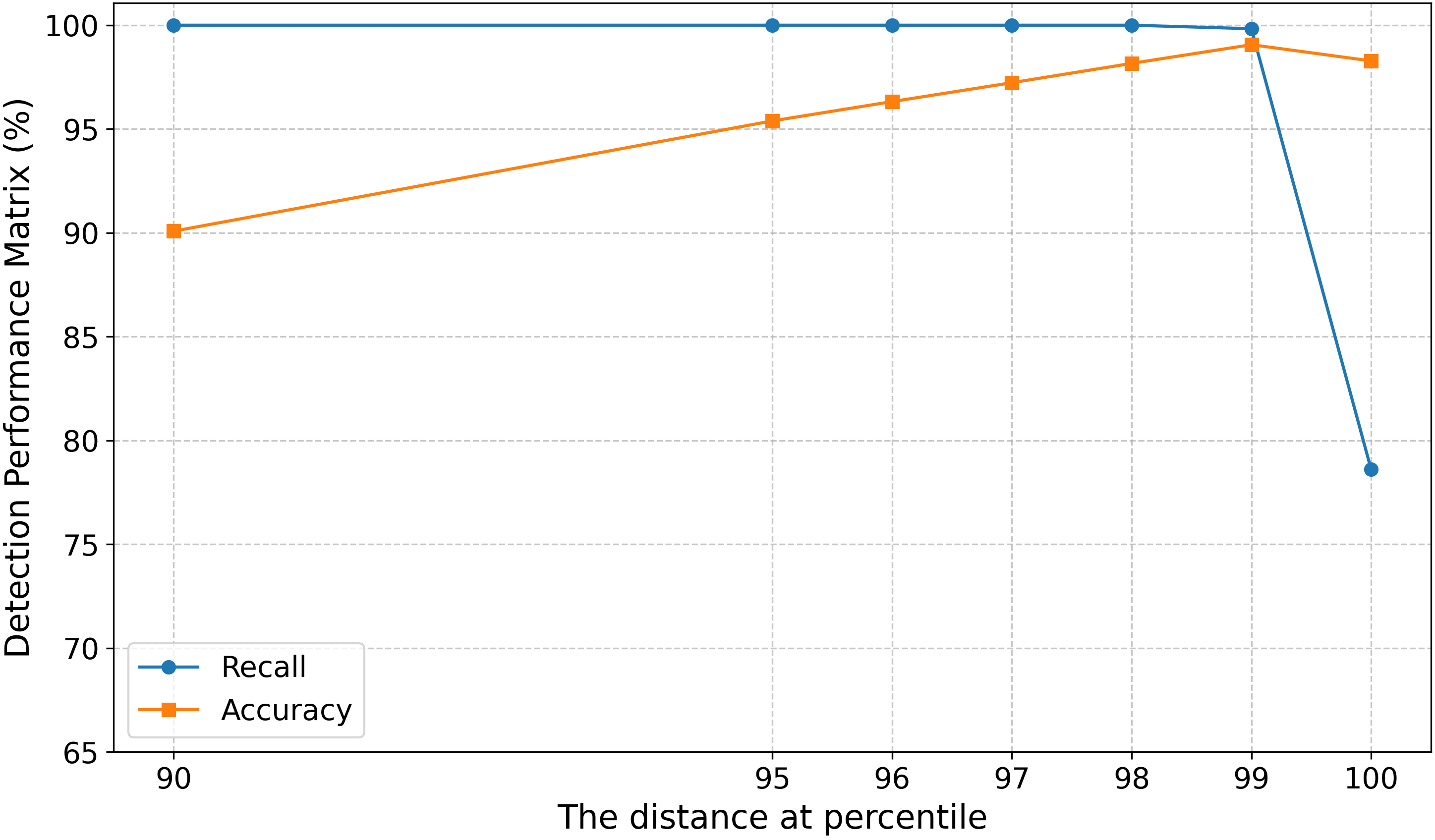}}
\caption{Detection performance of max threshold under different percentile.}
\label{scale_result}
\end{figure}

For detection, we want to identify poisoned samples as much as possible, even if we mistakenly classify a few clean samples as poisoned. So we want to adjust the max threshold, that is, not use the max Mahalanobis distance in $D_{bc}$ as the threshold. Instead, use the distance of a specific percentile of Clean samples (such as 95\%, the value corresponding to 95\% of the Mahalanobis distances of all samples in the dataset $D_{bc}$ in ascending order) as the threshold. Figure \ref{scale_result} further explores the scalability of the max threshold when dynamically adjusting the distance based on percentiles. At lower percentiles (90\%-98\%), the recall remains perfect (100\%), while the accuracy increases steadily from 90.08\% to 99.06\%. However, at the 100th percentile, the recall drops sharply to 78.6\%, while the accuracy drops slightly to 98.28\%. These results show that percentile-based thresholds provide flexibility but require careful tuning to effectively balance recall and accuracy. A recall of 100\% indicates that all poisoned samples can be identified, so we want to pursue higher accuracy at the highest possible recall.

Experimental results show that the proposed defense mechanism can effectively detect poisoned samples in semantic communication systems and maintain high accuracy and recall under different poisoning ratios. A key observation is that when the max threshold is set to percentile = 98\%, perfect detection of poisoned samples (100\%) is achieved in all different poisoning ratios, and the best accuracy is achieved when recall=100\%. When percentile = 99\%, nearly 100\% detection of poisoned samples is achieved, while better accuracy is achieved than percentile = 98\%. Overall, the defense mechanism achieves best performance and an optimal balance between recall and accuracy when the percentile is set to 98


\section{Conclusion} \label{Conclusion}
Semantic communication systems represent a transformative advancement in modern communication networks, but their reliance on GAI models (such as autoencoder) makes them vulnerable to Backdoor attacks. This paper proposed a defense mechanism that leverages semantic similarity analysis to detect Backdoor attacks in such systems. The key contributions of this paper include the development of two threshold-setting strategies, max similarity thresholds and mean similarity thresholds scaled by a factor, to detect Backdoor attack. Comprehensive experiments are conducted under various poisoning ratios and achieved high detection accuracy and recall. The results confirm the capability of the defense mechanism to distinguish poisoned samples reliably.

Future work will focus on extending the proposed defense mechanism to handle more complex data types, such as audio and video. Investigating the impact of diverse trigger patterns and more dynamic threshold-setting methods also remains an avenue for future research to strengthen system security against evolving Backdoor attack.






\bibliographystyle{unsrt}
\bibliography{ref}

\end{document}